\documentclass[aps,prx,amsmath,twocolumn,superscriptaddress,letterpaper,floatfix]{revtex4}
\usepackage{graphicx,color}
\usepackage{verbatim}
\usepackage{amssymb}   
\usepackage{amsmath}
\usepackage{amsfonts}
\usepackage{mathdots}
\usepackage{hyperref}
\usepackage{epsfig}

\usepackage{braket}
\usepackage{bm}
\begin{document}
	\title{Berry curvature, spin Hall effect and nonlinear optical response in moir\'e transition metal dichalcogenide heterobilayers}
	
	\author{Jin-Xin Hu}
	\author{Ying-Ming Xie}\thanks{Corresponding author: yxieai@connect.ust.hk}
	\author{K. T. Law} \thanks{Corresponding author: phlaw@ust.hk}
	
	\affiliation{Department of Physics, Hong Kong University of Science and Technology, Clear Water Bay, Hong Kong, China} 	
	
	\begin{abstract}
		Recently, topological flat bands and the spin Hall effect have been experimentally observed in AB-stacked MoTe$_2$/WSe$_2$ heterostructures. In this work, we systematically study the Berry curvature effects in moir\'{e} transition metal dichalcogenide (TMD) heterobilayers. We point out that the moir\'{e} potential of the remote conduction bands would induce a sizable periodic pseudo-magnetic field (PMF) on the valence band. This periodic PMF creates net Berry curvature flux in each valley of the moir\'{e} Brillouin zone. The combination of the effect of the Berry curvature and the spin-valley locking can induce the spin Hall effect being observed in the experiment. Interestingly, the valley-contrasting Berry curvature distribution generated by the PMF can be probed through shift currents, which are DC currents induced by linearly polarized lights through nonlinear responses. Our work sheds light on the novel quantum phenomena induced by Berry curvatures in moir\'e TMD heterobilayers.
	\end{abstract}
	\maketitle
	
	\section{\bf{Introduction}}
	The discovery of two-dimensional moir\'{e} materials leads to the engineering of new platforms for the study of novel topological, superconducting, and magnetic properties of electrons  in recent years  \cite{cao2018unconventional,cao2018correlated,yankowitz2019tuning,sharpe2019emergent,po2018origin,koshino2018maximally,serlin2020intrinsic,zhang2019twisted,balents2020superconductivity,mak2022semiconductor}. For example, magneto-electric and nonlinear Hall effects have been demonstrated in twisted graphene superlattice \cite{he2020giant,zhang2022giant,sinha2022berry,chakraborty2022nonlinear,pantaleon2021tunable} and twisted transition metal dichalcogenide (TMD) homobilayers \cite{hu2022nonlinear}.
	
	Notably, moir\'{e} TMD heterobilayers, in which moir\'e pattern mainly originated from the lattice mismatching between two distinct TMD layers, have been observed to exhibit nontrivial topological and correlated properties \cite{seyler2019signatures,tong2017topological,zhai2020layer,naik2022intralayer,zhang2022correlated,li2021imaging,wu2019topological,wang2020correlated,kang2022switchable,xu2022tunable,tao2022valley}. The study showed that a quantum anomalous Hall state at filling with $\nu=1$ (one hole per moir\'e unit cell) was observed in AB stacked moir\'{e} MoTe$_2$/WSe$_2$ heterobilayers \cite{li2021quantum,xie2022valley,xie2022topological,zhang2021spin,devakul2022quantum,pan2022topological,chang2022quantum,dong2023excitonic}. Very recently, the spin Hall torque has been demonstrated near $\nu=1$ and $\nu=2$ stemming from the large Berry curvature in this AB-stacked 2L-MoTe$_2$/WSe$_2$ heterostructures \cite{tschirhart2022intrinsic}. However, unlike the graphene moir\'e superlattice or twisted TMD homobilayers, the novel responses induced by the Berry curvature in TMD heterobilayers remain unknown theoretically. Moreover, in previous works \cite{wu2018hubbard,zhang2020moire,zhang2021electronic,angeli2021gamma}, the model for TMD heterobilayers is simply described by $H=-\bm{\hat{p}^2}/(2m)+V(\bm{r})$, where $\bm{\hat{p}}$ is the crystal momentum operator, $m$ is an electron effective mass and $V(\bm{r})$ is the moir\'{e} potential. As $H$ simply represents a valence band free Fermion moving in a periodic potential, the discovery of Berry curvature induced spin Hall effect in the experiment is quite surprising.
	
	In this work, we describe the  moir\'{e} TMD heterobilayers as a massive Dirac Fermion moving in a periodic moir\'e potential, in which the moir\'e potential of both conduction band and valence band is taken into account. Given that the low energy states are near the valence band edge, we project out the freedom of the conduction band by using the quantum commutation relation of crystal momentum $\hat{\bm{p}}$ and position $\hat{\bm{r}}$. Remarkably, we find that the moir\'e potential on the conduction band, which although being $1\sim 2$ eV away, contributes a periodic pseudo-magnetic field (PMF) to the valence band in the low energy state. We next show that the periodic PMF results in a  moir\'e valley-contrasting Berry curvature distribution, which exhibits net Berry curvature flux in each valley. Being consistent with the experiment in \cite{tschirhart2022intrinsic}, we find a large spin Hall effect in this case. It arises from a combination of the giant Ising spin-orbit coupling and the net Berry curvature flux induced by PMF. Finally, we show that the predicted moir\'e valley-contrasting Berry curvature distribution induced by the periodic PMF could exhibit a salient feature in the shift current response, which is a second-order DC response by applying a linear polarized light. The shift current response is tied to the quantum geometric properties of the system and varies microscopically due to changes in properties of the Bloch wavefunction upon excitation between bands \cite{sipe2000second,cook2017design,morimoto2016topological}. Due to the presence of valley-contrasting Berry curvature distribution, we find that the photocurrent as a function of photon energy exhibits two peaks and the peak separation is proportional to the strength of PMF. Our theory highlights that the periodic PMF plays an important role in the novel responses induced by Berry curvature in moir\'e heterobilayer TMDs.
	
	\section{\bf{Model Hamiltonian}}
\begin{figure}
		\centering
		\includegraphics[width=1\linewidth]{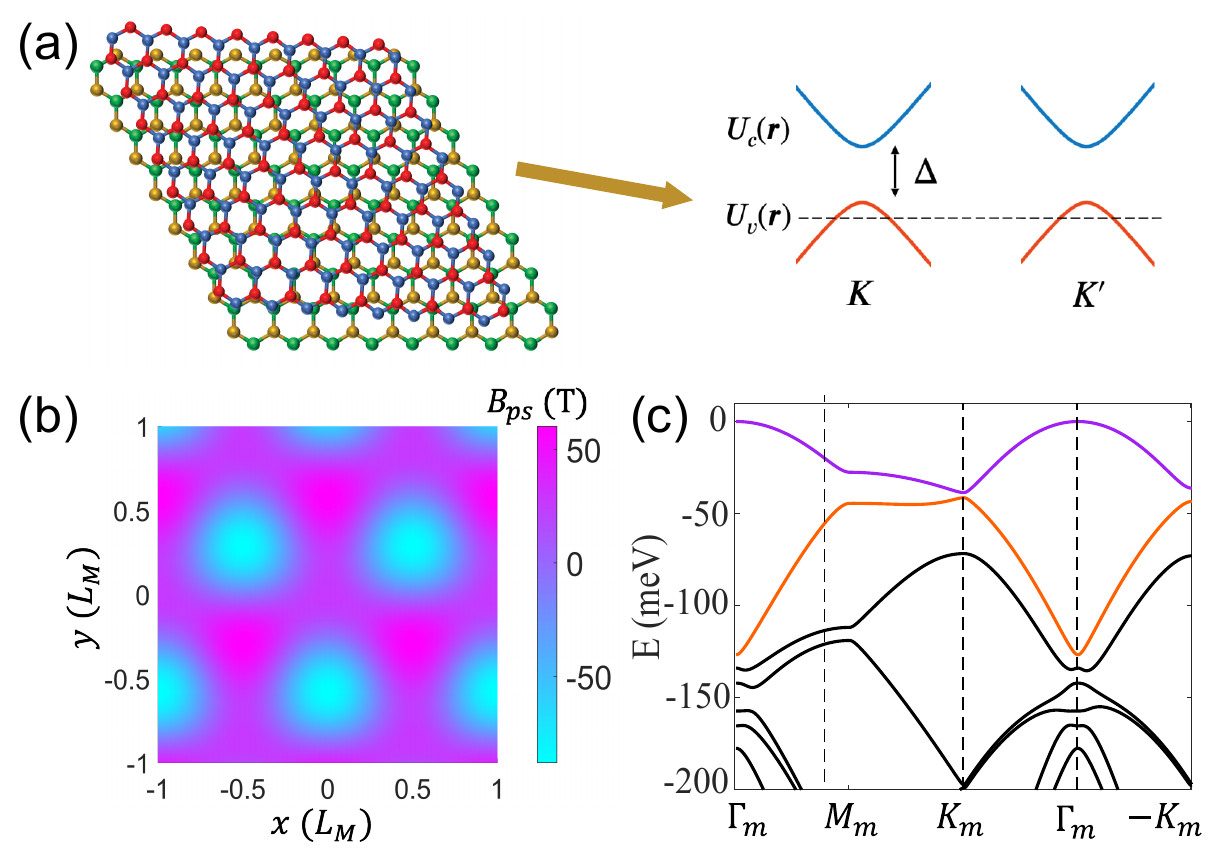}
		\caption{(a) The schematic picture of TMD heterobilayers, with a top layer (red and blue atoms) and bottom layer (yellow and green atoms). The low energy physics of the top layer is described by a massive Dirac model with a modified moir\'{e} potential. (b) The landscape of a $C_3$ symmetric periodic PMF indicated by Eq.(4). We set $U_c=20$ meV, $\phi_c=0.4$ $\pi$, $B_0=30$ T. (c) The calculated moir\'{e} bands with $B_0=$30 T, $\phi_c=0.4$ $\pi$, $U_v=12$ meV, $\phi_v=0.3$ $\pi$. The zero energy is shifted to the band edge.}
		\label{fig:fig1}
	\end{figure}

	Due to a large band offset (hundreds of meV) between the two layers in moir\'{e} TMD heterobilayers, we assume that the low energy states are arisen from one layer, while the other layer contributes to a periodic moir\'{e} potential. It is known that the 2H-TMD monolayer is described by massive Dirac Fermions \cite{xiao2012coupled}. For MoTe$_2$/WSe$_2$ heterobilayers, the valence band maximum of MoTe$_2$ is about 300 meV higher than WSe$_2$ \cite{yamaoka2018efficient}. We thus model the MoTe$_2$ layer with a massive Dirac Hamiltonian including slow-varying moir\'{e} potential on both conduction and valence band
	\begin{equation}
		\label{Eq1}
		H= v_F\left(
		\begin{matrix}{}
			&0 &\pi^\dagger \\
			&\pi &0
		\end{matrix}\right)+
		\left(
		\begin{matrix}{}
			&U_c(\bm{\hat{r}}) &0 \\
			&0 &U_v(\bm{\hat{r}})
		\end{matrix}\right)
		+\frac{\Delta}{2}\sigma_z,
	\end{equation}
	where $\hat{\pi}$ is the momentum operator with $\hat{\pi}=\tau \hat{p}_x+i\hat{p}_y$, $v_F$ is the Fermi velocity, $\Delta$ is the energy gap between the conduction band and the valence band, $\tau=\pm$ denote $K$ and $K'$ valleys. See Fig.~\ref{fig:fig1}(a) for an illustration of this model. $U_c$ and $U_v$ represent the moir\'{e} potential of conduction and valence band with $U_c(\bm{r})=2U_c\sum_{i=1}^{3}\cos(\bm{G}_i\cdot\bm{r}+\phi_c)$, $U_v(\bm{r})=2U_v\sum_{i=1}^{3}\cos(\bm{G}_i\cdot\bm{r}+\phi_v)$, which is dedicated by the $D_3$ point group symmetry. $\bm{G}_{j}=G_0(\sin(\frac{4(j-1)\pi}{3}),\cos(\frac{4(j-1)\pi}{3}))$, $G_0=\frac{4\pi}{\sqrt{3}L_{M}}$. To be specific,  we set the moir\'{e} lattice constant $L_M\approx 5$ nm, $v_F=4\times  10^5$ m/s and $\Delta=$1 eV, which are estimated from the MoTe$_2$/WSe$_2$ moir\'{e} heterobilayers \cite{su2022massive}.
	
	We next project out the conduction band and obtain a low energy effective Hamiltonian to describe the states near the valence band edge in moir\'e TMD heterobilayers. To the first order, we get the effective Hamiltonian
	\begin{equation}
		\label{Eq2}	H_{eff}=-\frac{1}{2m^*}\hat{\pi}(1-\frac{U_c(\bm{\hat{r}})}{\Delta})\hat{\pi}^\dagger+U_v(\bm{\hat{r}})-\frac{\Delta}{2},
	\end{equation}
	where $m^*$ is the effective mass with $m^*=\Delta/(2v_F^2)$. By using the commutation relation $[\bm{\hat{r}},\bm{\hat{p}}]=i\hbar$, we find the effective Hamiltonian becomes
	\begin{equation}
		\label{Eq3}
		H_{eff}=-\frac{1}{2m^*}(p_x^2+p_y^2+2e\tau\bm{p}\cdot\bm{A})+U_v(\bm{r})-\frac{\Delta}{2},
	\end{equation}
	where the vector potential $\bm{A}(\bm{r})=-A_0[\bm{a}_2\sin(\bm{G}_1\cdot\bm{r}+\phi_c)-\bm{a}_1\sin(\bm{G}_2\cdot\bm{r}+\phi_c)-\bm{a}_3\sin(\bm{G}_3\cdot\bm{r}+\phi_c)]$ with $A_0=\frac{\hbar U_c G_0}{e\Delta}$, $\bm{a}_1=(1/2,-\sqrt{3}/2),\bm{a}_2=(1,0),\bm{a}_3=\bm{a}_2-\bm{a}_1$. The vector potential $\bm{A}(\bm{r})$ obeys Coulomb gauge $\nabla\cdot \bm{A}(\bm{r})=0$. The details of deriving the continuum model are shown in Appendix A.
	
	Notably, we find besides the kinetic energy part, the effective Hamiltonian includes a $\bm{p}\cdot\bm{A}$ term. This term arises from the conduction band's moir\'{e} potential and the momentum-dependent mixing induced by the momentum operator $\hat{\pi}$. One can regard $\bm{A}$ as a gauge potential so that we define the PMF $B_{ps}(\bm{r})$ as
	\begin{equation}
		\label{Eq4}
		B_{ps}(\bm{r})=\partial_x A_y-\partial_y A_x=\tau B_0\sum_{i=1}^{3}\cos(\bm{G}_i\cdot\bm{r}+\phi_c),
	\end{equation}
	with the strength of PMF $B_0=\hbar U_c G_0^2/(e\Delta)$. The strength of PMF is mainly determined by the energy gap $\Delta$ and the conduction band moir\'e potential $U_c$. It is worth noting that the moir\'{e} potential on the valence band has no influence on the PMF though it plays an important role in the band structure.
	
	The topography of this PMF $B_{ps}(\bm{r})$ is shown in Fig.\ref{fig:fig1}~(b), which displays the same period as the moir\'{e} superlattice. By using a conduction band moir\'{e} potential $U_c=20$ meV and energy gap $\Delta=1$ eV, we find the PMF strength $B_0$ is as sizable as 30 T. Naively, it seems one can completely neglect the conduction band and its moir\'{e} potential as $\Delta$ is very large in this case. However, our finding points out that the conduction band's moir\'{e} potential would enable the states at the valence band to experience an effective PMF.
	
\begin{figure}
		\centering
		\includegraphics[width=1\linewidth]{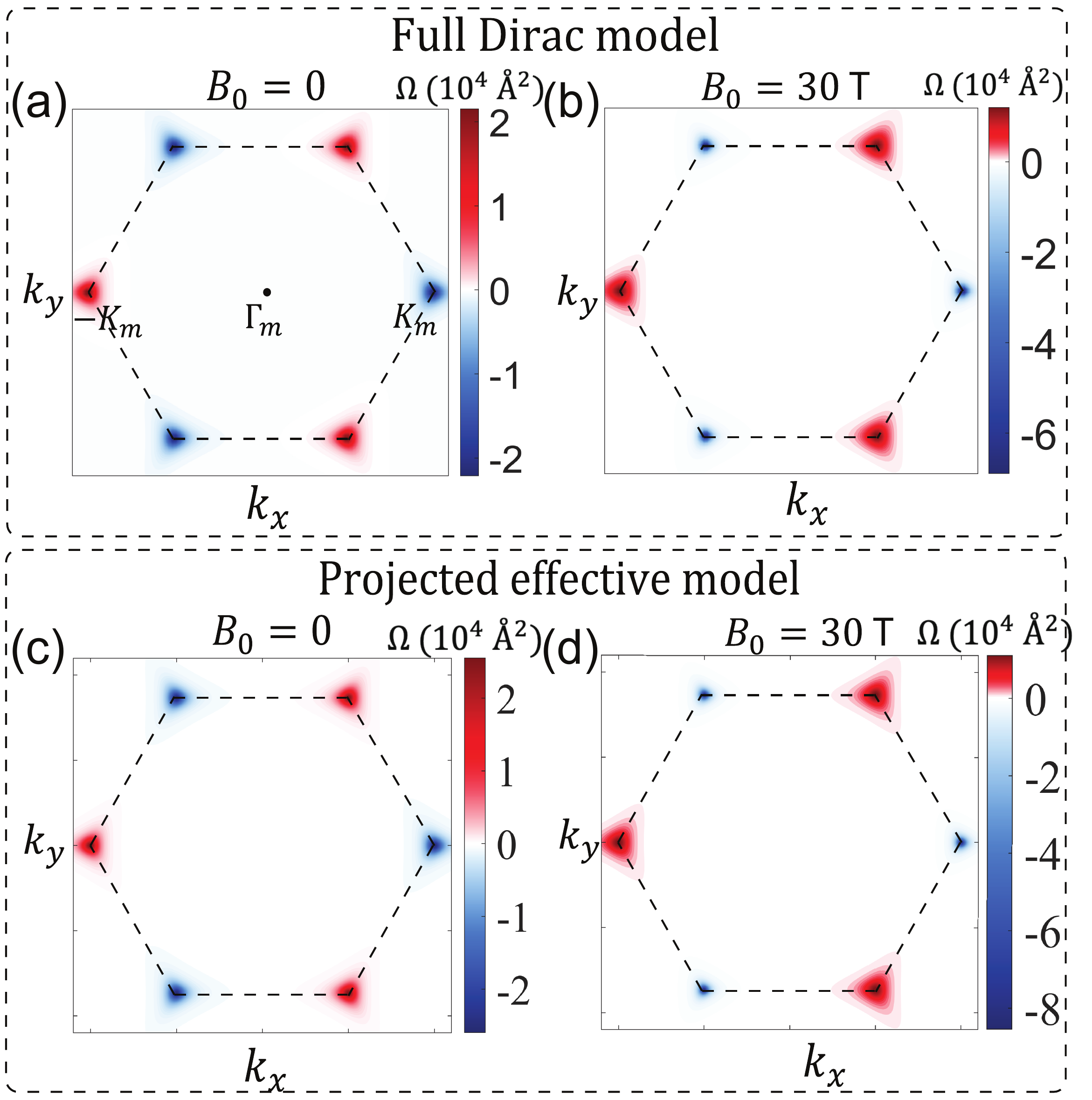}
		\caption{(a), (b) The Berry curvature $\Omega$ of the top moir\'{e} band with (a) $B_0=0$ and (b) $B_0=30$ T calculated by the full Dirac Hamiltonian in Eq.\ref{Eq1}. (c), (d) The Berry curvature $\Omega$ of the top moir\'{e} band with (c) $B_0=0$ and (d) $B_0=30$ T calculated by the projected effective Hamiltonian in Eq.\ref{Eq3}. The other parameters are set by $\phi_c=0.4$ $\pi$, $U_v=12$ meV and $\phi_v=0.3$ $\pi$.}
		\label{fig:fig2}
	\end{figure}

	To see how the PMF affects the moir\'{e} band structure of the TMD heterobilayers, we then diagonalize the effective Hamiltonian with plane wave basis. The resulting moir\'{e} bands of $K$-valley are plotted in Fig.\ref{fig:fig1}~(c), whereas the $K'$-valley is related by the time-reversal symmetry operation. To verify the accuracy of our projected effective continuum model, in Fig.\ref{fig:fig2} we compare the calculated Berry curvature of the top moir\'{e} band (purple band in Fig.\ref{fig:fig1}~(c)) between the full Dirac Hamiltonian in Eq.\ref{Eq1} (Fig.\ref{fig:fig2}~(a),(b)) and the effective Hamiltonian in Eq.\ref{Eq3} (Fig.\ref{fig:fig2}~(c),(d)), which shows a good agreement. It can be seen that there is a Berry curvature centering around $K_m$ and $-K_m$ pockets within the moir\'{e} Brillouin zone. The PMF enables a distinct gap between these two pockets.
	
\begin{figure}
		\centering
		\includegraphics[width=1\linewidth]{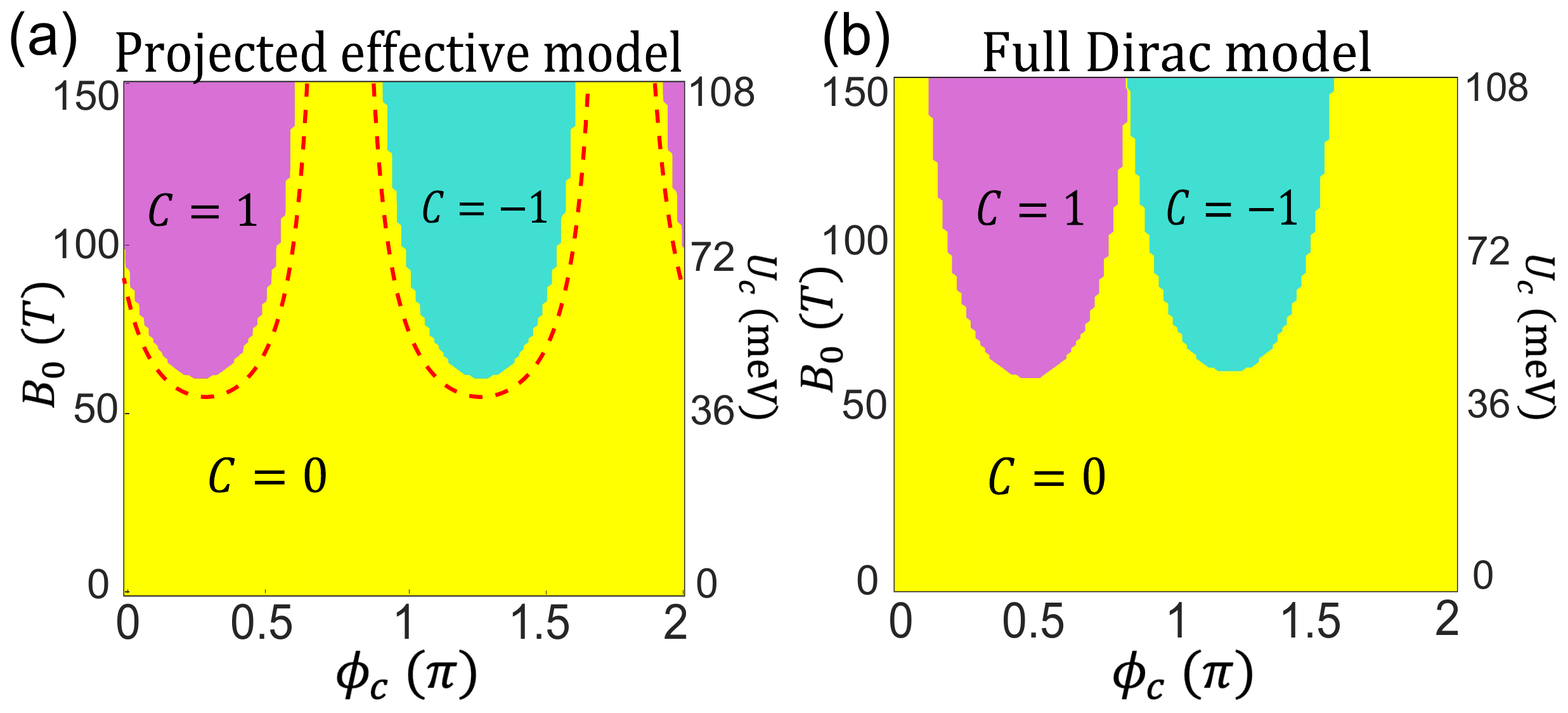}
		\caption{(a), (b) The topological phase diagram calculated by (a) the projected effective Hamiltonian in Eq.\ref{Eq3}, (b) the full Dirac Hamiltonian in Eq.\ref{Eq1} as a function of $B_0$ and $\phi_c$. The yellow (purple, blue) regions are the trivial (topological) phase with valley Chern number $C=0$ ($C=\pm 1$). The red dashed lines represent the phase boundaries given by Eq.(5).}
		\label{fig:fig3}
	\end{figure}

	By further tuning the conduction band's moir\'{e} potential $U_c$ to change the PMF, the top two moir\'{e} bands can further exchange Berry curvature by gap closing and reopening and undergo a topological phase transition. Following Ref.~\cite{xie2022valley} using the three-band continuum model near $\pm K_m$ point, we can obtain the topological phase transition boundary lines analytically
	\begin{equation}
		\label{Eq5}
		B_0\sin(\phi_c+\frac{\pi}{6})=\pm\frac{4\sqrt{3}m^*}{e\hbar}U_v\cos(\phi_v+\frac{\pi}{6}).
	\end{equation}
	In Fig.\ref{fig:fig3}~(a) we numerically calculate the topological phase diagram with various $B_0$ and $\phi_c$ by using the continuum model of Eq.\ref{Eq3}. To highlight the effect of PMF, we fix the moir\'{e} potential $U_v=12$ meV and $\phi_v=0.3$ $\pi$ throughout the main text. The phase transition boundary lines described by Eq.\ref{Eq5} are plotted as red dashed lines. In Fig.\ref{fig:fig3}~(b) we numerically calculate the topological phase diagram by using the full Dirac Hamiltonian of Eq.\ref{Eq1}, which basically agrees with Fig.\ref{fig:fig3}~(a). The error comes from the lost efficacy of first-order perturbation for large $U_c$. This provides the evidence that the mechanism of topological nature in moir\'{e} massive Dirac model is also from PMF \cite{su2022massive}. The details about the three-band continuum model near $\pm K_m$ point in this work are presented in Appendix B.
	
	It has also been proposed that a non-uniform strain distribution imposed on moir\'{e} TMD heterobilayers arising from lattice relaxation can also induce a topological phase transition \cite{xie2022valley}. The physical origin of the topology is shown to be understood in terms of the Haldane model with zero magnetic flux in a single unit cell \cite{haldane1988model}. The low energy Hamiltonian adopted in \cite{xie2022valley} is almost the same as Eq.\ref{Eq3}, but the origin of the PMF term arises from lattice relaxation. In this work, we point out an intrinsic origin to generate the PMF with conduction band moir\'{e} potential in common TMD heterobilayers. Clearly, a large PMF ($>50$ T) is needed to drive the system into topological regions, which is not to be realized in real systems readily. Thus the topological regions should be narrow and harsh. In the following sections, we will study the Berry curvature effects in the region with a low field (small $B_0$).
	
	\section{\bf{Spin Hall effect}}
	Apart from the nontrivial topology, the question is whether the valley-contrasting Berry curvature would induce some novel responses, which could help to identify the PMF effects in moir\'e TMD heterobilayers. In this section, we propose that a large spin Hall effect could be induced by the PMF, which may provide a plausible explanation for the spin Hall torque seen in MoTe$_2$/WSe$_2$ heterostructures recently \cite{tschirhart2022intrinsic}. The spin Hall effect appears when a spin current is generated perpendicularly to an electrical current. Because of the spin-valley locking and Ising spin-orbital coupling, the spin Hall effect is also  a valley Hall effect \cite{peng2020intrinsic}.
	
\begin{figure}
		\centering
		\includegraphics[width=1\linewidth]{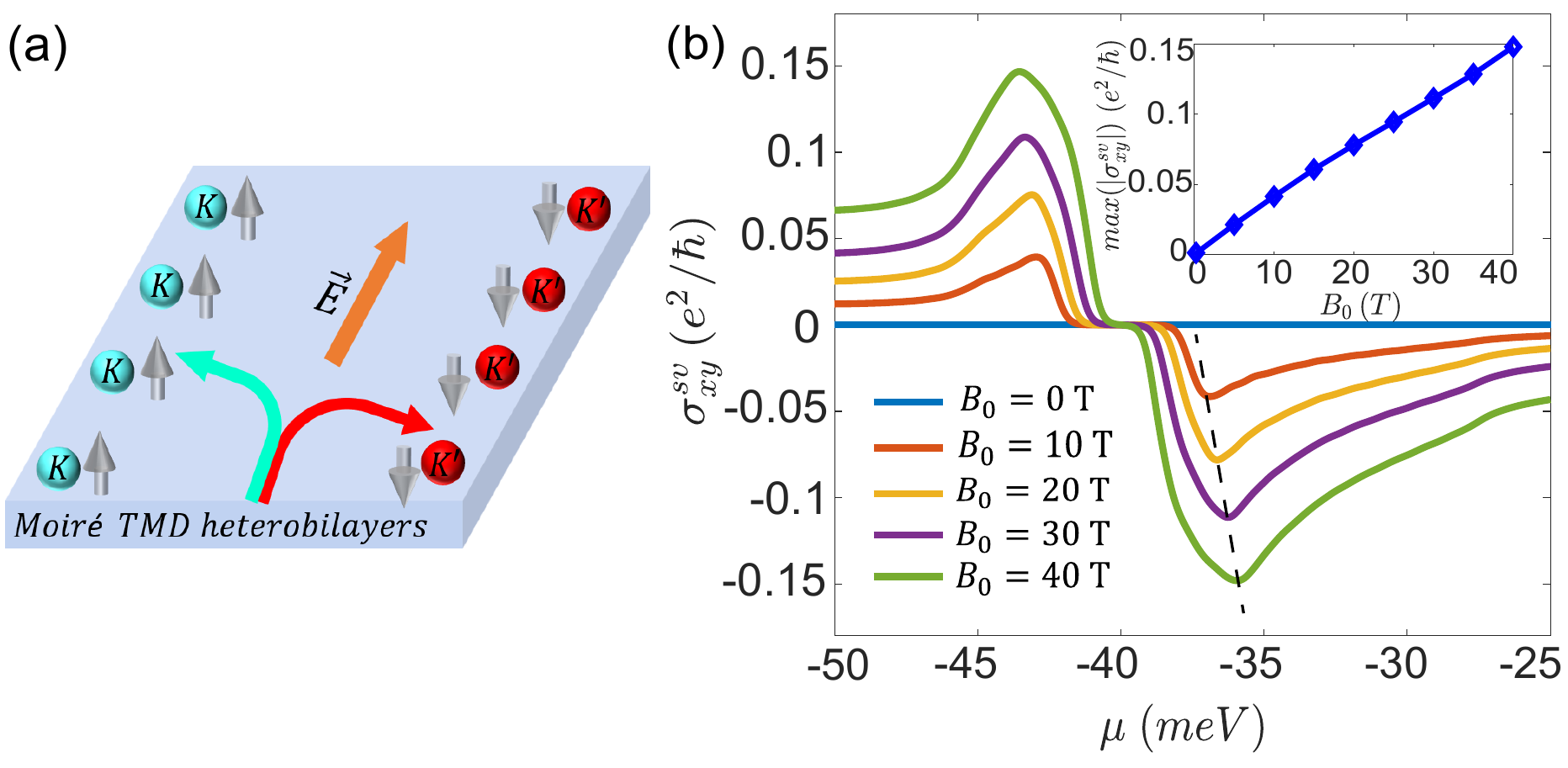}
		\caption{(a) A schematic picture of the spin valley Hall effect in the moir\'{e} TMD heterobilayers. The orange arrow indicates the in-plane current direction and the white arrows indicate the out-of-plane spin direction. $K$ and $K'$ indicate two valleys. (b) The calculated spin-valley Hall conductivity $\sigma_{xy}^{sv}$ as a function of a chemical potential $\mu$ ($\mu\approx$ -40 meV near the gap of top two moir\'e bands in Fig.\ref{fig:fig1}~(c)) with different strength of $B_0$. The temperature T is set to be 2 K. The inside panel is the dependence of the maximum value of $\sigma_{xy}^{sv}$ with $B_0$ (connecting with dashed line). We set $U_v=12$ meV,$\phi_v=0.3$ $\pi$, $\phi_c=0.4$ $\pi$.}
		\label{fig:fig4}
	\end{figure}

	Using the effective Hamiltonian in Eq.\ref{Eq3}, we can calculate the spin-valley Hall conductivity $\sigma_{xy}^{sv}$
	\begin{equation}
		\label{Eq6}
		\sigma_{xy}^{sv}=\frac{2e^2}{\hbar}\int \frac{d^2\bm{k}}{(2\pi)^2}[f_1(\bm{k})\Omega_1(\bm{k})+f_2(\bm{k})\Omega_2(\bm{k})],
	\end{equation}
	where 1 (2) is the band index of the first (second) moir\'{e} band in Fig.\ref{fig:fig1}~(c), $\Omega_{n}(\bm{k})$ is the  Berry curvature of $n$-th band, the integral is calculated over the moir\'{e} Brillouin zone, and $f_{1,2}(\bm{k})=\{1+\text{exp}[(E_{1,2}(\bm{k})-\mu)/k_B T]\}^{-1}$ are the Fermi-Dirac functions. Note that $\Omega$ is valley-contrasting due to the time-reversal symmetry ($\Omega_{1,2}^{\tau=+}=-\Omega_{1,2}^{\tau=-}$). As a result, under an in-plane electric field, $\Omega$ can drive charge carriers at opposite valleys to flow in opposite transverse directions, which leads to transverse spin-valley currents (Fig.\ref{fig:fig4}~(a)).
	
	In Fig.\ref{fig:fig4}~(b) we show the spin-valley Hall conductivity $\sigma_{xy}^{sv}$ for different $B_0$. For $B_0=0$, $\sigma_{xy}^{sv}=0$ because the spinless time reversal symmetry enforces $\Omega_{\bm{k}}=-\Omega_{-\bm{k}}$. In contrast, $\sigma_{xy}^{sv}$ becomes finite in the presence of the PMF. It is clear that $\sigma_{xy}^{sv}$ increases as the PMF strength $B_0$ increases, and the maximum value of $\sigma_{xy}^{sv}$ shows a linear increase at different values of $B_0$ which is shown in the inside panel. The order of $\sigma_{xy}^{sv}$ is about $0.1e^2/\hbar$, which is much larger than that in the monolayer TMD ($\sim$ 0.01$e^2/\hbar$) \cite{zhou2019spin}.
	
	To understand the monotonic increase of $\sigma_{xy}^{sv}$ as a function of $B_0$, we derive the Berry curvature near $\pm K_m$ points from the effective Hamiltonian (Appendix B): $\Omega^{s,l}_{\bm{k}}=-sl v_F^2 m_0^{s}/[8(m_0^{s 2}+v_F^2 k^2/4)^{3/2}]$ with $s=\pm1$ for $\pm K_m$ pockets, and $l=\pm1$ for band index ($+1$ for the upper band and $-1$ for the lower band, $m_0^{s}$ is the effective mass of $s$ pocket). By integrating over $\pm K_m$ pockets, we can evaluate the spin-valley Hall conductivity  $\sigma_{xy}^{sv}$ at zero temperature analytically according the Eq.~(\ref{Eq6}),
	\begin{equation}
		\label{Eq7}
		\sigma_{xy}^{sv}=\left\{
		\begin{array}{rcl}
			0, & & {0\leq |\mu|<m_0^+}\\
			-\text{sgn}(\mu)\frac{e^2}{2\pi\hbar}(1-\frac{m_0^+}{|\mu|}), & & {m_0^+\leq|\mu|<m_0^-}\\
			-\text{sgn}(\mu)\frac{e^2}{2\pi\hbar}\frac{m_0^--m_0^+}{|\mu|}, & & {|\mu|\geq m_0^-},
		\end{array}\right.
	\end{equation}
	where $m_0^{\pm}=\sqrt{3}U_v\cos(\phi_v+\frac{\pi}{6}) \mp \frac{\sqrt{3}}{2}g\sin(\phi_c+\frac{\pi}{6})$ with $g=\hbar eB_0/(2\sqrt{3}m^*)$. Indeed, we find
	\begin{equation}
		\label{Eq8}
		\text{max}(|\sigma_{xy}^{sv}|) \approx \frac{e^2}{2\pi\hbar}\frac{g\sin(\phi_c+\pi/6)}{U_v\cos(\phi_v+\frac{\pi}{6})},
	\end{equation}
	which indicates $\text{max}(|\sigma_{xy}^{sv}|)$ is approximately linear with $B_0$. The underlying reason is  that the increasing of PMF strength enable a larger net Berry curvature flux in each valley (see Fig.\ref{fig:fig2}~(d)).
	
	Therefore,  we have demonstrated that in spite of the large gap between conduction and valence band in a massive Dirac model, the PMF on the valence band is generated by a moir\'{e} modulation. Such PMF would enable the presence of a large spin Hall effect induced by the valley contrasting Berry curvature between the top two moir\'{e} bands in moir\'{e} TMD heterobilayers.
	
	\section{\bf{Terahertz optical responses}}
\begin{figure}
		\centering
		\includegraphics[width=1\linewidth]{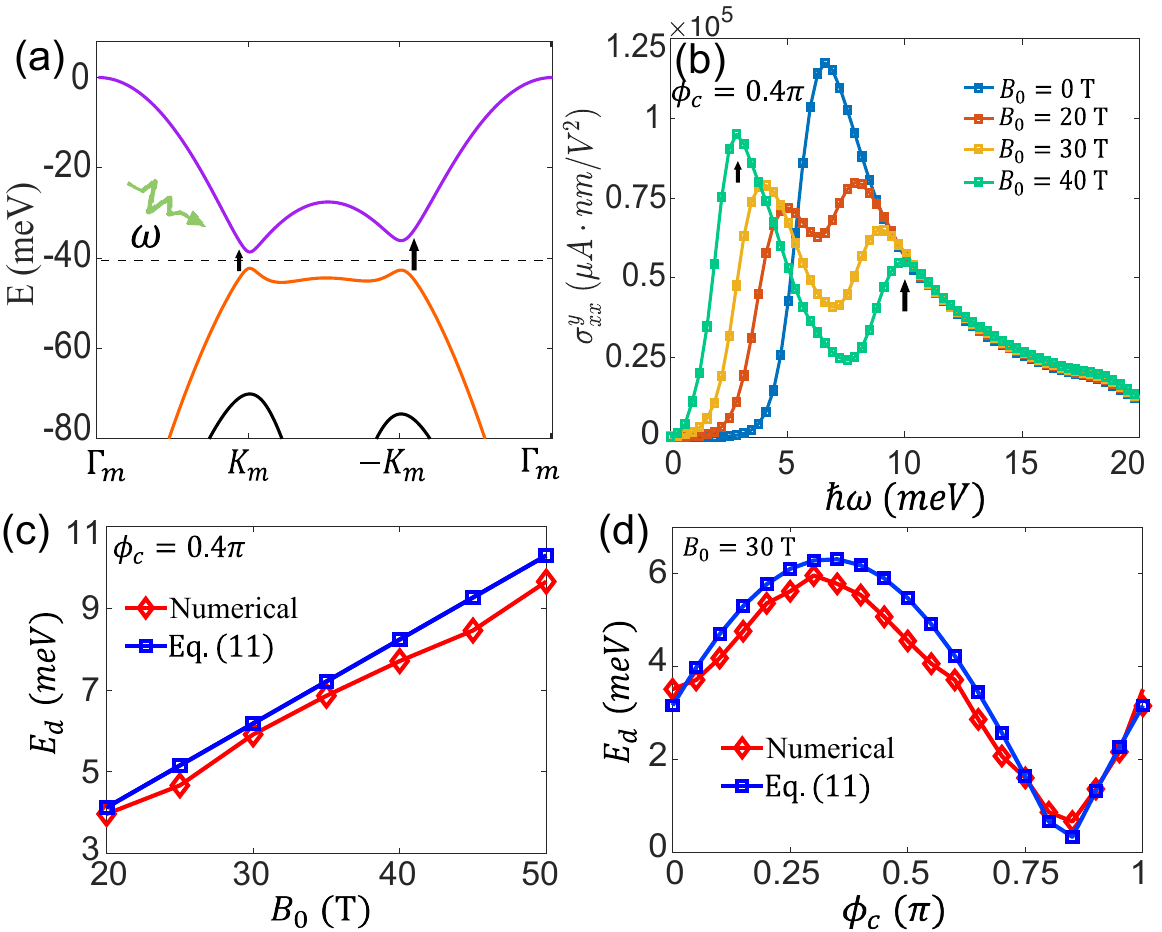}
		\caption{(a) The moir\'{e} band structure with $B_0=30$ T, $\phi_c=0.4$ $\pi$. The arrows represent the interband optical transitions. The dashed line labels the position of the chemical potential. (b) The shift current photoconductivity for different $B_0$ from the optical transitions in (a). (c) and (d) are the comparison of peak difference which shows the numerical calculation and theoretical calculation in Eq.\ref{Eq11}. For example, the peak difference for $B_0=40$ T is depicted by the photon energy difference (indicated by black arrows).}
		\label{fig:fig5}
	\end{figure}

	As we have shown in the previous section, the PMF would influence the Berry curvature effects of TMD heterobilayers significantly. Next, we show the PMF strength can be explicitly observed in the experiment by studying the terahertz optical responses of  TMD heterobilayers. We set the chemical potential near $\nu=2$ (two holes per moir\'{e} unit cell) so that the relevant states contributing to the terahertz response would contain the information of the PMF (see Fig.\ref{fig:fig5}~(a)).
	
	Before presenting the results of nonlinear terahertz optical responses, we actually first looked at linear optical conductivity $\sigma_{\alpha\beta}(\omega)$, where $\alpha, \beta$ labels the polarized direction of the light.  We find the longitudinal optical conductivity $\sigma_{\alpha\alpha}$ is almost insensitive to the PMF, because the value of $\sigma_{\alpha\alpha}$ mainly reflects the inter-band linear resonant optical response strength while the Berry curvature is not that essential in this case. Interestingly, we find that the spin-valley optical conductivity defined as $\sigma_{xy}^{sv}(\omega)=\sigma^{\tau=+}_{xy} (\omega)-\sigma^{\tau=-}_{xy}(\omega)$ can be enhanced by the PMF. However, we still find that in general, it is hard to intuitively see the strength of PMF from the linear optical response only. More details about the linear optical conductivity of this system are presented in Appendix C.
	
	According to the previous works \cite{chaudhary2022shift,morimoto2016topological}, nonlinear terahertz optical responses can reflect the topological nature of wavefunctions. On the other hand, we have shown the PMF can induce a valley-contrasting Berry curvature. To manifest the PMF strength through optical responses, we thus now look at the second-order nonlinear terahertz optical response. As we will show that the shift current response can fit our purpose, which measures a DC photocurrent driven in second-order optical response in noncentrosymmetric quantum materials by shining a linear polarized light.
	
	The shift current characterizes the nontrivial band topology of the moir\'{e} bands in the optical transition process. With a electric field $E_{\beta}(\omega)$ at frequency $\omega$ and linearly polarized in the $\beta$ direction, the shift current $J$ in the $\alpha$ direction takes the form
	\begin{equation}
		\label{Eq9}
		J_{\alpha}=\sigma_{\beta\beta}^\alpha E_{\beta}(\omega)E_{\beta}(-\omega),
	\end{equation}
	where the second-order conductivity tensor $\sigma_{\beta\beta}^\alpha$ has the form \cite{cook2017design}
	\begin{equation}
		\label{Eq10}
		\sigma_{\beta\beta}^\alpha(\omega)=\frac{2g_{s}\pi e^3}{\hbar^2 S}\sum_{nm,\bm{k}} f_{nm} \text{Im}(r_{mn}^\beta r_{nm;\alpha}^\beta)\delta(\omega_{nm}-\omega),
	\end{equation}
	where $S$ is the sample area, $g_s=2$ is the spin (valley) degeneracy, $n$ and $m$ are band indexes and $\omega$ is the photon frequency. The occupation difference $f_{nm}=f_n-f_m$ with $f_n$ being the Fermi-Dirac distribution of band $n$. $r_{mn}^\beta$ are the inter-band Berry connections defined as $r_{mn}^\beta=i\langle m|\partial_{k_\beta}|n\rangle$. And the generalized derivative  $r_{nm;\alpha}^\beta=\partial_{k_\alpha}r_{nm}^\beta-i(A_{nn}^\alpha-A_{mm}^\alpha)r_{nm}^\beta$, where $A_{nn}^\alpha=i\langle n|\partial_{k_\alpha}|n\rangle$ are intraband Berry connections for band $n$. The non-vanishing tensor  $\sigma_{\beta\beta}^\alpha(\omega)$ can be deduced from $D_3$ point group symmetry  generated by $C_{3z}$ and $C_{2y}$. According to the symmetry constraint of the $D_3$ point group, the non-zero elements in shift current optical conductivity tensor are $\sigma_{xx}^y=\sigma_{xy}^x=\sigma_{yx}^x=-\sigma_{yy}^y$. Without loss of generality, we display the results with $\sigma_{xx}^y$ in the following.
	
	Figure.\ref{fig:fig5}~(b) shows the photon energy dependence of the shift current photoconductivity $\sigma_{xx}^y (\omega)$ at different PMF strength $B_0$, where the Fermi energy is in the gap between the first and the second moir\'{e} bands. We note that (i) the order of photoconductivity is $\sim 10^5$ $\mu\text{A}\cdot nm/\text{V}^2$, which is very large and is in the same order as the one in twist bilayer graphene \cite{chaudhary2022shift}; (ii) the photoconductivity curve develops two peaks and their separation increases with  the PMF strength.
	
	The two peaks stem from the concentration of Berry curvature near $K_m$ and $-K_m$ pockets. The photon energy difference of the two peaks reflects the opposite shifting of the Dirac mass by PMF  at $K_m$ and $-K_m$ pockets. In other words, the separation of two peaks can be estimated by the gap difference at $K$ and $-K$ points, which we denote as $E_d$. $E_d$ can be also solved from the three-band continuum model near $\pm K_m$ as well, which gives
	\begin{equation}
		\label{Eq11}
		E_d= \frac{\hbar eB_0}{ m^*} |\sin(\phi_c+\frac{\pi}{6})|.	
	\end{equation}
	
	In Fig.\ref{fig:fig5}~(c) and (d), we compare the peak to peak frequency difference $E_d$ between the numerical result (from continuum model) and theoretical calculation (in Eq.~(\ref{Eq11})), which shows a good agreement. Fig.\ref{fig:fig5}~(c) and (d) show the $B_0$ and $\phi_c$ dependence of $E_d$, respectively. The $E_d$ is monochromatically linear with $B_0$ and periodic with $\phi_c$, where $E_d$ increases to $9$ meV when $B_0=50$ T with $\phi_c=0.4$ $\pi$.
	The peak separation on shift current photoconductivity curve in principle is resolvable in a terahertz optical measurement. Moreover, the peak intensity at higher $\hbar\omega$ decreases and has a redshift, while the lower energy peak has a blueshift and the intensity increases with $B_0$. This is because the interband Berry connection $r_{mn}$ gets enhanced when the gap at $K_m$ pocket gets smaller (Fig.\ref{fig:fig5}~a).

	\section{\bf{conclusion}}
	In a conclusion, we have studied the Berry curvature effects in  heterobilayer TMD superlattice in this work. In particular, we have found that  the periodic PMF  plays a crucial role in affecting the Berry curvature distribution of moir\'e bands. Importantly, we found that  the conduction band moir\'e potential within a massive Dirac Hamiltonian naturally induces a periodic PMF upon the valence band.  We have also pointed out  how the large spin Hall effect observed in the experiment could be explained by the moir\'e valley-contrasting Berry curvature distribution induced by the PMF. In our model, $U_c$, $\phi_c$, $U_v$, and $\phi_v$ are parameters that are determined by DFT calculations and experimental conditions and there are no mutual constraint relationships between them.
	
	Furthermore, we have demonstrated the observation of a two-peak splitting in shift current photoconductivity would provide direct evidence of periodic PMF in TMD heterobilayers. Our theoretical findings in this work are general, which can be verified via transport and optical measurements in various recent fabricated TMD heterobilayers, such as MoTe$_2$/WSe$_2$\cite{tschirhart2022intrinsic}, MoSe$_2$/WSe$_2$\cite{baek2020highly} and MoS$_2$/WSe$_2$\cite{rivera2018interlayer}.
	
	\section*{\bf{ACKNOWLEDGEMENTS}}
	We thank Cheng-Ping Zhang for inspiring discussions. K.T.L. acknowledges the support of the Ministry of Science and Technology, China and HKRGC through 2020YFA0309600, RFS2021-6S03, C6025-19G, AoE/P-701/20, 16310520, 16310219, 16307622 and 16309718. Y.M.X. acknowledges the support of HKRGC through PDFS2223-6S01.

	\appendix
	\renewcommand{\theequation}{A-\arabic{equation}}
	\renewcommand\thefigure{A-\arabic{figure}}
	\setcounter{equation}{0}
	\setcounter{figure}{0}
	\section{DERIVATION OF THE CONTINUUM MODEL}\label{AppendixA}

	In this section, we give the derivation of the model Hamiltonian in detail. We start from a massive Dirac model including moir\'{e} potential same as Eq.\ref{Eq1} in main text,
	\begin{equation}
		\label{EqA1}
		H=v_F(\tau \hat{p}_x\sigma_x+\hat{p}_y\sigma_y)+\frac{\Delta}{2}\sigma_z+
		\left(
		\begin{matrix}{}
			&U_c(\bm{\hat{r}}) &0 \\
			&0 &U_v(\bm{\hat{r}})
		\end{matrix}\right),
	\end{equation}
	where $U_c$, $U_v$ represent the moir\'{e} potential of conduction and valence band with $U_c(\bm{r})=2U_c\sum_{i=1}^{3}\cos(\bm{G}_i\cdot\bm{r}+\phi_c)$, $U_v(\bm{r})=2U_v\sum_{i=1}^{3}\cos(\bm{G}_i\cdot\bm{r}+\phi_v)$. $\bm{G}_1=(0,1)G_0,\bm{G}_2=(-\sqrt{3}/2,-1/2)G_0,\bm{G}_3=(\sqrt{3}/2,-1/2)G_0, G_0=\frac{4\pi}{\sqrt{3}L_M}$.\\
	By using the two spinor wavefunction $(\Psi_c,\Psi_v)^T$, the Schr\"{o}dinger equation can be written in the form of two
	coupled equations
	\begin{eqnarray}
		\label{EqA2} (\frac{\Delta}{2}+U_c(\bm{\hat{r}}))\Psi_c+v_F(\tau \hat{p}_x-i\hat{p}_y)\Psi_v &=& E\Psi_c, \\
		\label{EqA3} v_F(\tau \hat{p}_x+i\hat{p}_y)\Psi_c-(\frac{\Delta}{2}-U_v(\bm{\hat{r}}))\Psi_v &=& E\Psi_v.
	\end{eqnarray}
	Since the energy gap $\Delta$ is relatively large, we can do the approximation that $E\approx -\Delta/2$ after considering the states near the valence band edge. Thus from Eq.\ref{EqA2} we obtain
	\begin{equation}
		\label{EqA4}
		\Psi_c=-\frac{v_F}{\Delta+U_c(\bm{\hat{r}})}(\tau \hat{p}_x-i\hat{p}_y)\Psi_v.
	\end{equation}
	Insert Eq.\ref{EqA4} into Eq.\ref{EqA3}, and we obtain
	\begin{equation}
		\label{EqA5}
		[-v_F^2(\tau \hat{p}_x+i\hat{p}_y)\frac{1}{\Delta+U_c(\bm{\hat{r}})}(\tau \hat{p}_x-i\hat{p}_y)-\frac{\Delta}{2}+U_v(\bm{\hat{r}})]\Psi_v=E\Psi_v.
	\end{equation}
	By expanding $\frac{1}{\Delta+U_c(\bm{\hat{r}})}$ to the first order, we get the effective Hamiltonian
	\begin{equation}
		\label{EqA6}
		H_{eff}=-\frac{v_F^2}{\Delta}\hat{\pi}(1-\frac{U_c(\bm{\hat{r}})}{\Delta})\hat{\pi}^\dagger-\frac{\Delta}{2}+U_v(\bm{\hat{r}})
	\end{equation}
	with the momentum operator $\hat{\pi}=\tau \hat{p}_x+i\hat{p}_y$. To deal with the term $\hat{\pi}U_c(\bm{\hat{r}})\hat{\pi}^\dagger$, we first divide it into a self-hermitian operator
	\begin{equation}
		\label{EqA7}
		\begin{split}
			\hat{\pi}{U_c}(\bm{\hat{r}})\hat{\pi}^\dagger &=1/2*([\hat{\pi},{U_c}(\bm{\hat{r}})]\hat{\pi}^\dagger+\hat{\pi}[{U_c}(\bm{\hat{r}}),\hat{\pi}^\dagger]\\
			&+{U_c}(\bm{\hat{r}})\hat{p}^2+\hat{p}^2{U_c}(\bm{\hat{r}}))
		\end{split}
	\end{equation}
	with $\hat{p}^2=\hat{p}_x^2+\hat{p}_y^2$.
	The commutation relation is $[\hat{\pi},U_c(\bm{\hat{r}})]=\hbar(-i\tau\partial_x+\partial_y)U_c(\bm{\hat{r}})$. Using the plane waves $|\bm{k}\rangle=e^{i\bm{k}\cdot \bm{r}}$, we can obtain $\hat{\pi}|\bm{k}\rangle=(\tau p_x+ip_y)|\bm{k}\rangle$ and $\hat{\pi}^\dagger|\bm{k}\rangle=(\tau p_x-ip_y)|\bm{k}\rangle$.
	
	Thus we can write the effective continuum model
	\begin{equation}
		\label{EqA8}
		\begin{split}
			H_{eff}&=-\frac{v_F^2}{\Delta}\pi(1-\frac{U_c(\bm{r})}{\Delta})\pi^\dagger-\frac{\Delta}{2}+U_v(\bm{r})\\
			&=-\frac{v_F^2}{\Delta}[(1-\frac{U_c(\bm{r})}{\Delta})(p_x^2+p_y^2)+2e\bm{p}\cdot\bm{A}]+U_v(\bm{r})-\frac{\Delta}{2}\\
			&\approx -\frac{v_F^2}{\Delta}(p_x^2+p_y^2+2e\bm{p}\cdot \bm{A})+U_v(\bm{r})-\frac{\Delta}{2},
		\end{split}
	\end{equation}
	where the charge $e$ is to make the dimension of $\bm{A}$ to be the gauge potential.
	The vector potential $\bm{A}$ satisfies
	\begin{align}
		\begin{split}
			\label{EqA9}A_x&=-\tau A_0[\sin(\bm{G}_1\cdot\bm{r}+\phi_c)-\frac{1}{2}\sin(\bm{G}_2\cdot\bm{r}+\phi_c)\\
			&-\frac{1}{2}\sin(\bm{G}_3\cdot\bm{r}+\phi_c)]
		\end{split}\\
		\label{EqA10}A_y&=-\tau A_0[\frac{\sqrt{3}}{2}\sin(\bm{G}_2\cdot\bm{r}+\phi_c)-\frac{\sqrt{3}}{2}\sin(\bm{G}_3\cdot\bm{r}+\phi_c)]
	\end{align}
	where $A_0=\frac{\hbar U_c G_0}{e\Delta}$.
	
\begin{figure}
		\centering
		\includegraphics[width=1\linewidth]{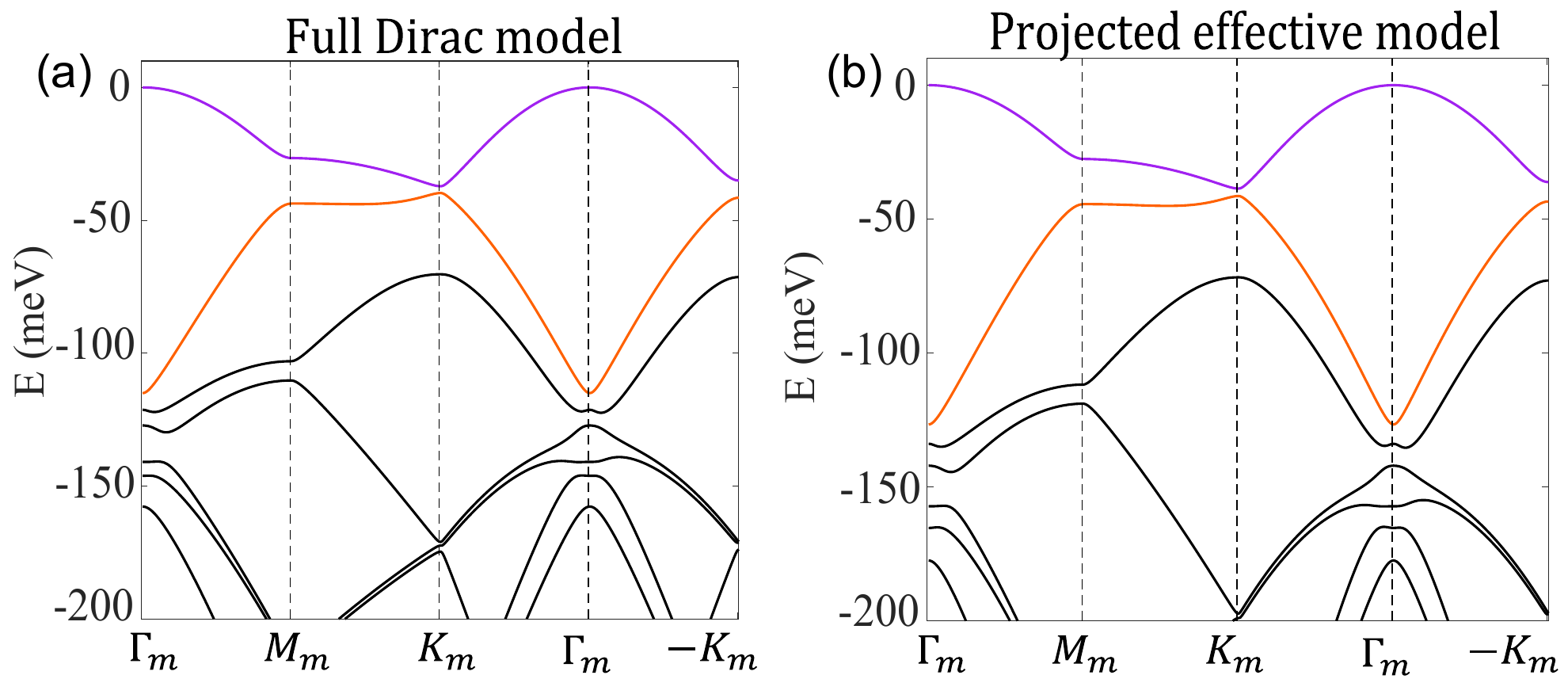}
		\caption{(a) The band structure calculated by full Dirac model in Eq.\ref{EqA1}. (b) The band structure calculated by effective continuum model in Eq.\ref{EqA8}.}
		\label{fig:figS1}
	\end{figure}

	In Eq.~(\ref{EqA8}), we find besides the kinetic energy part, the effective Hamiltonian includes a $\bm{p}\cdot\bm{A}$ like term induced by a pseudo-magnetic field. The pseudo-magnetic field is given by
	\begin{equation}
		\label{EqA11}
		B(\bm{r})=\tau\frac{\hbar U_c G_0^2}{e\Delta}\sum_{i=1}^{3}\cos(\bm{G}_i\cdot\bm{r}+\phi_c).
	\end{equation}
	This result reveals that the moir\'{e} potential of the conduction band triggers a gauge potential on the valence band, with opposite signs in two valleys. In Fig.\ref{fig:figS1} we compare the band structures calculated by the full Dirac model (Eq.~(\ref{EqA1})) and projected effective continuum model (Eq.~(\ref{EqA8})). The results show that the effective continuum model works well.
	
	\renewcommand{\theequation}{B-\arabic{equation}}
	\renewcommand\thefigure{B-\arabic{figure}}
	\setcounter{equation}{0}
	\setcounter{figure}{0}
	\section{DERIVATION OF THREE-BAND EFFECTIVE HAMILTONIAN}\label{AppendixB}
	\begin{figure}
		\centering
		\includegraphics[width=1\linewidth]{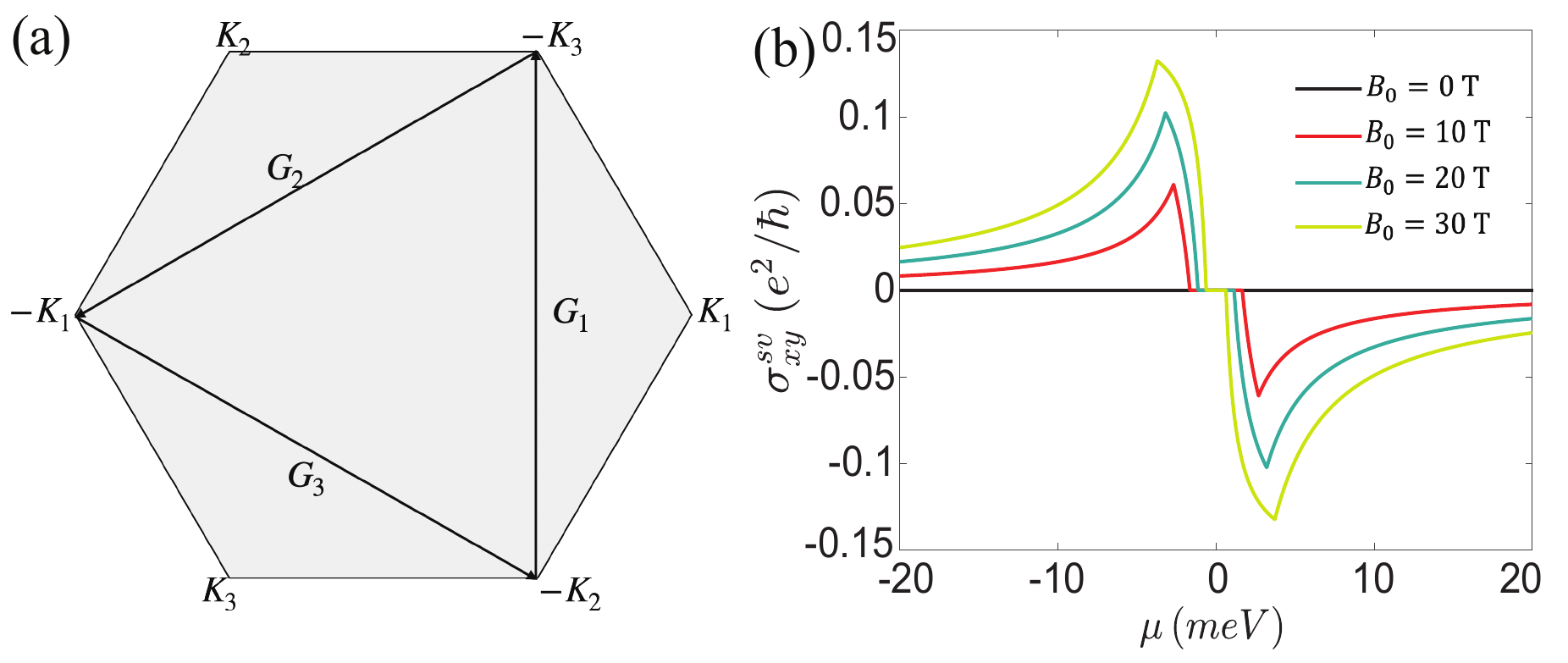}
		\caption{(a) A sketch of the moir\'{e} Brillouin zone and how the moir\'{e} Brillouin corners $K$ are connected by reciprocal lattice vectors $G_j$. (b) Theoretical calculation of $\sigma_{xy}^{sv}$ as a function of $\mu$ with $B_0=0$, 10 T,20 T, 30 T.}
		\label{fig:figS2}
	\end{figure}
	In this Appendix section, we derive the three-band effective model from continuum model Eq.~(\ref{EqA8}) at the Brillouin corners (in Fig.\ref{fig:figS2}~(a)). First we consider the case of $U_c(\bm{r})=0$ and $U_v(\bm{r})=0$. Because the three corners of moir\'{e} Brillouin are connected by the superlattice reciprocal vectors, using the plane waves $|\bm{k}\rangle=e^{i\bm{k}\cdot \bm{r}}$, the effective Hamiltonian near $\pm \bm{K}$ is written as
	\begin{equation}
		\label{EqB1}
		H^0_{\pm}(\bm{k})=\epsilon_0 I\mp\left(
		\begin{matrix}{}
			&vk_x &0 &0 \\
			&0 & v(-\frac{1}{2}k_x+\frac{\sqrt{3}}{2}k_y) &0 \\
			&0 &0 &v(-\frac{1}{2}k_x-\frac{\sqrt{3}}{2}k_y)
		\end{matrix}\right),
	\end{equation}
	where $\epsilon_0=-\hbar^2|K|^2/(2m^*),v=\hbar^2|K|/m^*$.
	For the moir\'{e} potential $U_v(\bm{r})$,
	\begin{equation}
		\label{EqB2}
		H^U_{\pm}(\bm{k})=\left(
		\begin{matrix}{}
			&0 &U_ve^{\pm i\phi_v} &U_ve^{\mp i\phi_v}\\
			&U_ve^{\mp i\phi_v}&0 &U_ve^{\pm i\phi_v}\\
			&U_ve^{\pm i\phi_v}&U_v e^{\mp i\phi_v} &0
		\end{matrix}\right).
	\end{equation}
	For the gauge field term,
	\begin{equation}
		\label{EqB3}
		H^A_{\pm}(\bm{k})=\left(
		\begin{matrix}{}
			&0 &-\frac{g}{2i}e^{\pm i\phi_c} &\frac{g}{2i}e^{\mp i\phi_c}\\
			&\frac{g}{2i}e^{\mp i\phi_c}&0 &-\frac{g}{2i}e^{\pm i\phi_c}\\
			&-\frac{g}{2i}e^{\pm i\phi_c}&\frac{g}{2i}e^{\mp i\phi_c} &0
		\end{matrix}\right).
	\end{equation}
	with $g=\hbar eB_0/(2\sqrt{3}m^*)$. At the Brillouin zone corners, the eigenenergies and eigenfunctions of $H^U+H^A$ are
	
	\begin{equation}
		\label{EqB4}
		\begin{split}
			E_1&=2(U_v\cos\phi_v \mp \frac{g}{2}\sin\phi_c), \\
			|\psi_1 \rangle &= \frac{1}{\sqrt{3}}(|\pm K_1\rangle+|\pm K_2\rangle+|\pm K_1\rangle),
		\end{split}
	\end{equation}
	\begin{equation}
		\label{EqB5}
		\begin{split}
			E_2&=2U_v\cos(\phi_v+\frac{2\pi}{3}) \pm g\sin(\phi_c-\frac{\pi}{3}), \\
			|\psi_2 \rangle &=\frac{i}{\sqrt{3}}(|\pm K_1\rangle+e^{\pm i\frac{2\pi}{3}}|\pm K_2\rangle+e^{\mp i \frac{2\pi}{3}}|\pm K_3\rangle),
		\end{split}
	\end{equation}
	\begin{equation}\label{EqB6}
		\begin{split}
			E_3&=2U_v\cos(\phi_v-\frac{2\pi}{3}) \pm g\sin(\phi_c+\frac{\pi}{3}), \\
			|\psi_3 \rangle &=\frac{-i}{\sqrt{3}}(|\pm K_1\rangle+e^{\mp i\frac{2\pi}{3}}|\pm K_2\rangle+e^{\pm i \frac{2\pi}{3}}|\pm K_3\rangle).
		\end{split}
	\end{equation}
	Thus in the basis spanned by ($|\psi_1 \rangle,|\psi_2 \rangle,|\psi_3 \rangle$), we can write the effective model which describes the states near $\pm K$ of the first three moir\'{e} bands:
	\begin{widetext}
		\begin{equation}\label{EqB7}
			H^{eff}_{\pm}(\bm{k})=\left(
			\begin{matrix}{}
				&2(U_v\cos\phi_v\mp\frac{g}{2}\sin\phi_c) &\frac{1}{2}v(k_y \mp ik_x) &\frac{1}{2}v(k_y\pm ik_x)\\
				&\frac{1}{2}v(k_y \pm ik_x) &2U_v\cos(\phi_v+\frac{2\pi}{3}) \pm g\sin(\phi_c-\frac{\pi}{3}) &\pm\frac{1}{2}v(k_x\pm ik_y)\\
				&\frac{1}{2}v(k_y\mp ik_x) &\pm\frac{1}{2}v(k_x\mp ik_y) &2U_v\cos(\phi_v-\frac{2\pi}{3}) \pm g\sin(\phi_c+\frac{\pi}{3})
			\end{matrix}\right).
		\end{equation}
	\end{widetext}
	
	The energy gap difference between $K_m$ and $-K_m$ pockets can be evaluated as $E_d=2\sqrt{3}g\sin(\phi_c+\pi/6)$.
	To calculate the spin Hall effect, we can only consider the first two bands and it becomes a massive Dirac model with the Fermi velocity $v_0=v_F/2$ and the mass $m_0^{\pm}=\sqrt{3}U_v\cos(\phi_v+\frac{\pi}{6}) \mp \frac{\sqrt{3}}{2}g\sin(\phi_c+\frac{\pi}{6})$. By further tuning the conduction band moir\'{e} potential $U_c$ to change the PMF, the top two moir\'{e} bands can further exchange Berry curvature by gap closing and reopening and undergo a topological phase transition. The gap-closing lines which characterize the topological phase transition can be obtained as
	\begin{equation}\label{EqB8}
		g\sin(\phi_c+\frac{\pi}{6})=\pm2U_v\cos(\phi_v+\frac{\pi}{6}).
	\end{equation}
	The pseudomagnetic field can drive the system more easily when $U_v$ is small and $\phi_v$ is close to $\pi/3$. The topological phase diagram is shown in the main text.
	We can calculate the Berry curvature near $\pm K_m$ points:
	\begin{equation}\label{EqB9}
		\Omega^{s,l}_{\bm{k}}=-sl\frac{v_0^2 m_0^{s}}{2(m_0^{s 2}+v_0^2 k^2)^{3/2}}
	\end{equation}
	with $s=\pm1$ for $\pm K$ points, and $l=\pm1$ for band index ($+1$ for the upper band and $-1$ for the lower band).
	By integrating over $K_m$ and $-K_m$ pockets, we can obtain the spin valley Hall conductivity $\sigma_{xy}^{sv}$
	\begin{equation}\label{EqB10}
		\sigma_{xy}^{sv}=\left\{
		\begin{array}{rcl}
			0, & & {0\leq |\mu|<m_0^+}\\
			-\text{sgn}(\mu)\frac{e^2}{2\pi\hbar}(1-\frac{m_0^+}{|\mu|}), & & {m_0^+\leq|\mu|<m_0^-}\\
			-\text{sgn}(\mu)\frac{e^2}{2\pi\hbar}\frac{m_0^--m_0^+}{|\mu|}, & & {|\mu|\geq m_0^-},
		\end{array}\right.
	\end{equation}
	In Fig.\ref{fig:figS1}~(b) we plot the $\sigma_{xy}^{xv}$ as the function of $\mu$ at zero temperature. And we obtain:
	\begin{equation}\label{EqB11}
		\begin{split}
			max(|\sigma_{xy}^{sv}|)&=\frac{e^2}{2\pi\hbar}\frac{\sqrt{3}g\sin(\phi_c+\pi/6)}{\sqrt{3}U_v\cos(\phi_v+\frac{\pi}{6}) + \frac{\sqrt{3}}{2}g\sin(\phi_c+\frac{\pi}{6})}\\
			&\approx \frac{e^2}{2\pi\hbar}\frac{g\sin(\phi_c+\pi/6)}{U_v\cos(\phi_v+\frac{\pi}{6})},
		\end{split}
	\end{equation}
	which means in the low $B$ field region, $max(|\sigma_{xy}^{sv}|)\sim B_0 \sin(\phi_c+\pi/6)$. 
	
	\renewcommand{\theequation}{C-\arabic{equation}}
	\renewcommand\thefigure{C-\arabic{figure}}
	\setcounter{equation}{0}
	\setcounter{figure}{0}
	\section{LINEAR OPTICAL CONDUCTIVITY}\label{AppendixC}
	\begin{figure}
		\centering
		\includegraphics[width=1.0\linewidth]{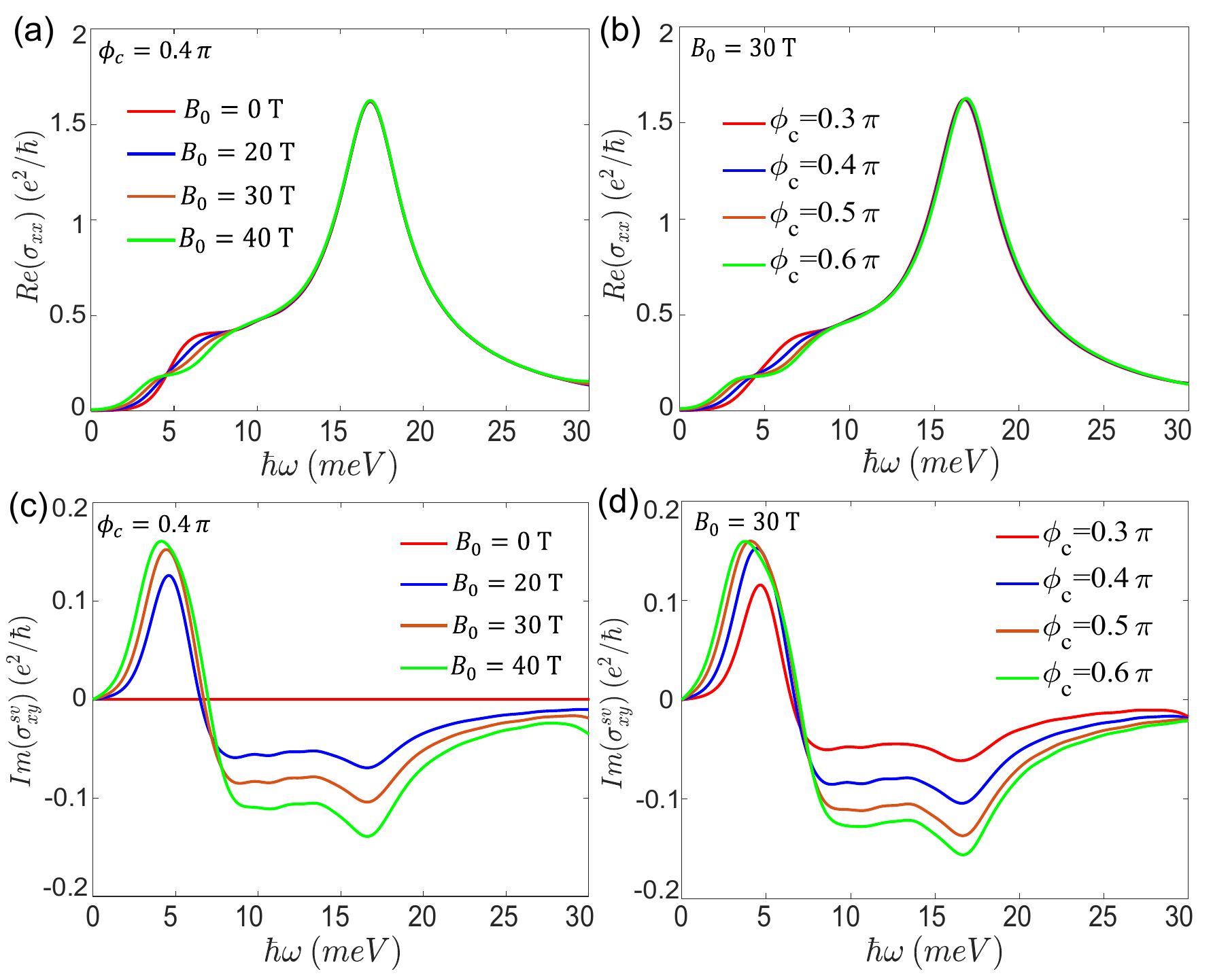}
		\caption{(a) ((c)) Real(imaginary) part of the longitudinal (transverse) optical conductivity with $B_0=0,20,30,40$ T and $\phi_c=0.4$ $\pi$, $U_v=12$ meV, $\phi_v=0.3$ $\pi$. (b) ((d))Real(imaginary) part of the longitudinal (transverse) optical conductivity with $\phi_c=0.3$ $\pi$, 0.4 $\pi$, 0.5 $\pi$, 0.6 $\pi$ and $B_0=30$ T, $U_v=12$ meV, $\phi_v=0.3$ $\pi$.}
		\label{fig:figS3}
	\end{figure}
	For a circularly polarized light, the optical conductivity is written in terms of the longitudinal part $\sigma_{xx}$ and transverse part $\sigma_{xy}$
	\begin{equation}
		\label{Eqc1}
		\sigma_{\pm}(\omega)=\sigma_{xx}(\omega)\pm i\sigma_{xy}(\omega),
	\end{equation}
	where $\pm$ for left($+1$) or right($-1$) circular polarization. Thus the dissipative components of the conductivity tensor is Re$(\sigma_{xx})$ and Im$(\sigma_{xy})$. The optical conductivity from inter-band transition can be calculated using standard linear response theory \cite{novelli2020optical}
	\begin{equation}\label{Eqc2}
		\begin{split}
			\sigma_{\alpha\beta}(\omega)=&-i\frac{e^2}{\hbar}\sum_{m\neq n}\int \frac{d^2\bm{k}}{(2\pi)^2}\frac{f_{\bm{k},m}-f_{\bm{k},n}}{E_{\bm{k},m}-E_{\bm{k},n}}\\
			&\frac{\langle u_{\bm{k}m}|\hat{v}_\alpha(\bm{k})|u_{\bm{k}n}\rangle \langle u_{\bm{k}n}|\hat{v}_\beta(\bm{k})|u_{\bm{k}m}\rangle}{\hbar \omega +i\eta+E_{\bm{k},m}-E_{\bm{k},n}}.
		\end{split}
	\end{equation}
	Because of the time-reversal symmetry, the conductivity in the transverse direction is $\sigma^{\tau=1}_{xy}(\omega)=-\sigma^{\tau=-1}_{xy}(\omega)$. As a result, $\sigma_{xy}(\omega)$ is spin-resolved due to the opposite spin from two valleys. The optical spin-valley conductivity is then defined as $\sigma_{xy}^{sv}(\omega)=\sigma^{\tau=1}_{xy}(\omega)-\sigma^{\tau=-1}_{xy}(\omega)$.
	
	In Fig.\ref{fig:figS3}~(a) and (b) we plot the Re$(\sigma_{xx})$ for different $B_0$ and $\phi_c$, and we find it does not depend on $B_{ps}$, for the longitudinal part of optical conductivity has a peak at the energy which corresponds to the mean gap of the two bands near the Brillouin zone boundary. In Fig.\ref{fig:figS3}~(c),(d) we show our results for the $\sigma_{xy}^{sv}$ vs.$\hbar\omega$ in units of $e/\hbar$ for the four values of $B_0$ and find it zero when $B_0=0$. The spinless time-reversal symmetry enforces the integral in Eq.\ref{Eqc2} to be zero and the $\bm{k}\cdot \bm{A}$ term in the Hamiltonian will break the spinless time-reversal symmetry and results in finite optical conductivity. It is clear that Im$(\sigma_{xy}^{sv})$ gets enhanced as $B_0$ increases due to the breaking of the spinless time-reversal symmetry.

	\renewcommand{\theequation}{D-\arabic{equation}}
	\renewcommand\thefigure{D-\arabic{figure}}
	\setcounter{equation}{0}
	\setcounter{figure}{0}
	\section{SYMMETRY ANALYSIS of THE NONLINEAR OPTICAL RESPONSE}\label{AppendixD}
	
	In this section, we discuss the symmetry properties of the shift current conductivity tensor. In the matrix form, the shift current conductivity tensor is expressed as:
	\begin{equation}\label{EqD1}
		\hat{\sigma}_{ab}^c=\left(
		\begin{matrix}{}
			&\sigma_{xx}^x &\sigma_{xy}^x &\sigma_{yx}^x &\sigma_{yy}^x \\
			&\sigma_{xx}^y &\sigma_{xy}^y &\sigma_{yx}^y &\sigma_{yy}^y
		\end{matrix}\right).
	\end{equation}
	In a symmetry operator $g$, $\hat{\sigma}_{ab}^c$ is transformed as:
	\begin{equation}\label{EqD2}
		\hat{\sigma}_{ab}^c \longrightarrow \hat{U}^\dagger(g)\hat{\sigma}_{ab}^c[\hat{U}(g)\otimes \hat{U}(g)].
	\end{equation}
	Here $\hat{U}(g)$ denotes the symmetry operation representation of the group element $g$. In $C_3$ rotation, $\hat{U}(C_3)=e^{i\frac{2\pi}{3}\sigma_y}$, it enforces:
	\begin{eqnarray}
		\label{EqD3}\sigma_{yy}^x=\sigma_{yx}^y=\sigma_{xy}^y=-\sigma_{xx}^x\\
		\label{EqD4}\sigma_{xx}^y=\sigma_{xy}^x=\sigma_{yx}^x=-\sigma_{yy}^y.
	\end{eqnarray}
	Similarly, in $C_{2y}$ rotation symmetry, it enforces:
	\begin{equation}\label{EqD5}
		\sigma_{yy}^x=\sigma_{yx}^y=\sigma_{xy}^y=-\sigma_{xx}^x=0.
	\end{equation}
	Thus the only nonzero and nonequivalent term in $\hat{\sigma}_{ab}^c$ is $\sigma_{xx}^y$, which is calculated in the main text.

\end{document}